\documentclass[5p,times,twocolumn,left=1in, right=1in]{elsarticle}
\journal{Physica D}

\usepackage[utf8]{inputenc}
\usepackage{amsfonts} 
\usepackage{amsmath}
\usepackage{amssymb}
\usepackage[hyperref]{xcolor}
\usepackage{bm}
\usepackage[hyphens]{xurl}
\usepackage{hyperref}
    \hypersetup{colorlinks=true}

\begin{document}
\hypersetup{
  linkcolor=blue,
  urlcolor=blue,
  citecolor=blue
}
\title{Learning dissipation and instability fields from chaotic dynamics}




\author[LGT]{Ludovico T Giorgini\corref{cor1}}
\ead{ludogio@mit.edu}            
\affiliation[LGT]{
                 organization={Department of Mathematics},
                 addressline={MIT},
                 city={Cambridge},
                 state={MA},
                 country={USA}}
\cortext[cor1]{Corresponding author}

\author[AS]{Andre N Souza}
\ead{sandre@mit.edu}    
\ead[url]{sandreza.github.io}
\affiliation[AS]{
                 organization={Department of Earth, Atmospheric, and Planetary Sciences},
                 addressline={MIT},
                 city={Cambridge},
                 state={MA},
                 country={USA}}

\author[DL]{Domenico Lippolis}
\ead{domenico@ujs.edu.cn}
\affiliation[DL]{organization={
            School of Mathematical Sciences},
            addressline={Jiangsu University}, 
            city={Zhenjiang},
            postcode={212013}, 
            country={China},
            }

\author[PC]{Predrag Cvitanovi{\'c}}
\ead{predrag.cvitanovic@physics.gatech.edu}            
\ead[url]{ChaosBook.org}
\affiliation[PC]{organization={
            Center for Nonlinear Science, School of Physics},
            addressline={Georgia Institute of Technology}, 
            city={Atlanta},
            postcode={30332-0430}, 
            state={GA},
            country={USA},
            }

\author[PS]{Peter Schmid}
\ead{peter.schmid@kaust.edu.sa}
\affiliation[PS]{organization={
            Physical Science and Engineering Division},
            addressline={KAUST}, 
            city={Thuwal},
            postcode={23955-6900}, 
            country={Saudi Arabia},
            }


\begin{abstract}
To make predictions or design control, information on local sensitivity of initial conditions and state-space contraction is both central, and often instrumental. However, it is not always simple to reliably determine instability fields or local dissipation rates, due to computational challenges or ignorance of the governing equations.  Here, we construct an alternative route towards that goal, by
estimating the Jacobian of a discrete-time dynamical system locally from the entries of the transition matrix that approximates the Perron-Frobenius operator for 
a given state-space partition. Numerical tests on one- and two-dimensional chaotic maps show promising results.
\end{abstract}

\maketitle

\section{Introduction}

The task of reconstructing the governing equations of a dynamical system from observed data is a principal challenge across various fields of theoretical and applied physics. This endeavor is not only academically significant but also has practical implications in diverse scientific disciplines, where understanding the underlying mechanisms of observed phenomena is paramount \cite{lejarza2022data}. Even if the dynamical systems under consideration are exceedingly high-dimensional, making an accurate description of the interactions between all scales and variables unfeasible, low-order models can often be constructed to retain the most important statistical and dynamical features of the system, effectively capturing the essential behavior without the need for exhaustive detail \cite{Holmes2012, Brunton2016, Schmid2010, santos2021reduced}. These simplified models provide a practical means of reconstructing the core dynamics and allow for meaningful analysis and prediction, even in complex, high-dimensional contexts \cite{Chorin2007, Moehlis2004, Kantz2004, giorgini2020precursors, Lim2020Predicting}.

Reduced-order models play a crucial role, for example, in climate physics by effectively describing interacting degrees of freedom and capturing key feedback mechanisms across spatial and temporal scales \cite{lorenz1963, Schneider2017, giorgini2022non, giorgini2021modeling, Palmer2000, Majda2012, keyes2023stochastic,baldovin2022extracting, vissio2018proof, giorgini202rresponse}. In finance, reduced-order models are used to understand market dynamics and to develop robust economic models, such as the Black-Scholes model \cite{black1973}, a key tool in financial mathematics. In biology, particularly in the study of ecosystems, disease propagation, and epidemiology, reduced-order models like the Lotka-Volterra equations provide insights into complex interactions, such as predator-prey dynamics \cite{lotka1925}. 

Despite remarkable success, 
reconstructing low-order models from observed data becomes particularly challenging in highly nonlinear and chaotic systems, where the inherent complexity and sensitivity to initial conditions make it difficult to accurately capture the essential dynamics using simplified models \cite{strogatz1994, cvitanovic2005chaos,vulpiani2010chaos}. Nonetheless, although a precise mathematical model driving observed data remains elusive, the statistical features of the dynamical system can often be robustly estimated from data, provided a statistically significant amount of observations is available \cite{souza2023statistical, geogdzhayev2024evolving, giorgini2023clustering, souza2023transforming, Souza_2024_2, souza2024modified}.
A notable example is given by the Koopman operator approach~\cite{Koopmanism, KoopLan}, where the generally chaotic, nonlinear dynamics of single state-space trajectories is modeled through a linear yet infinite-dimensional operator that maps observables forward. Its adjoint, the Perron-Frobenius operator, maps forward probability densities.
Relying on Markovianity, the Perron-Frobenius operator is approximated by a finite matrix, whose entries may be learned from the dynamics~\cite{KenBuhl}. Similar approaches have proven effective in the study of directed networks~\cite{DuccioDirNet}, and brain state and transitions in computational neuroscience~\cite{KringDeco}.   

In this work, we leverage the formal evolution of the Perron-Frobenius operator, together with the
established methods of approximation of the transport operator with a finite transfer matrix,  in order to
extract information on local dissipation and instability fields 
over the available state space. That is feasible through the
connection between the transfer matrix and the Jacobian
of the dynamical system under study.

We focus here on discrete-time dynamical systems of form $\bm{x}^{n+1} = \bm{f}(\bm{x}^n)$,
where $ n \in \mathbb{N}$ is the $n$th discrete time step, $ \bm{x} = \{x_1, \dots, x_D\} $ is a $D$-dimensional vector representing the state of the system, and $\bm{f}: \mathbb{R}^D \rightarrow \mathbb{R}^D$ is the forward map that characterizes the evolution of the system in time. For non-autonomous dynamical systems, we augment the dimension of the state vector, in order to incorporating external time-dependent changes.

The key observation is that a density of trajectories at time $n$, $\rho^{n}(\bm{x})$, is transformed by the Perron-Frobenius operator as \cite{CBmeasure}
\[
    \rho^{n+1}(\bm{x}) = \sum_{k=1}^{M} \frac{\rho^{n}(\bm{f}^{-1}_{k}(\bm{x}))}{|J(\bm{f}^{-1}_{k}(\bm{x}))|}\,.
\]
As explained below, in the evolution of a density by the Perron-Frobenius operator, which is in general a contraction \cite{Driebe}, one can identify the local dissipation with the Jacobian $J$ of the map,
computed on the $M$ preimages $\bm{f}^{-1}_{k}(\bm{x})$. We shall show that a lower bound for the
local Jacobian $J(\bm{x})$ may also be estimated from the entries of the transfer matrix. In one dimension, $J(\bm{x})$ is the derivative of the map.  

Knowledge of dissipation and of instability fields is essential for analysis, forecast, and control.
In particular, potential applications of our methodology include stabilizing an unstable system \cite{Ott1990}, enhancing the sensitivity to initial conditions in chaotic systems \cite{Wolf1985}, or ensuring that conservation laws are adhered to in physical models \cite{Thompson2002, Wiggins2003}, when direct evaluation of the Jacobians is unavailable due to computational challenges, or ignorance of the governing equations, as in experimental time series. This knowledge is invaluable not only for understanding of the intrinsic properties of the system, but also in facilitating the design of more effective and controlled dynamical models. This paper aims to demonstrate the utility of this approach through a detailed theoretical analysis and application to simulated data, highlighting its potential impact across a broad spectrum of scientific disciplines.

The article is structured as follows. Section \ref{sec2} explains how the Jacobian of the map  can be inferred from the transition matrix, in Section \ref{sec3} we apply this methodology to one- and two-dimensional chaotic maps, and Section \ref{conclusions} presents our conclusions.

\section{From the transition matrix to the Jacobian}
\label{sec2}

The \emph{Perron--Frobenius operator}  \(\mathcal{P}\) encapsulates how a probability density function \(\rho^n(\bm{x})\) at time step \(n\) evolves under the flow induced by the forward map $\bm{f}$,
\[
    \rho^{n+1}(\bm{x}) 
    \;=\; \bigl(\mathcal{P}\rho^n\bigr)(\bm{x})
\;.
\]
To construct a \emph{discrete} approximation of this operator, we partition the state space into \({N}\) equally sized \emph{control volumes}, 
\begin{equation}  
\bm{X} = [\bm{X}_1,\dots,\bm{X}_{N}]
\,.
    \label{contrVol}
\end{equation}
Each control volume has measure
\[
    \mu(\bm{X}_i) \;=\; \frac{\Omega}{N}, 
\]
where \(\Omega\) is the total volume of the explored region. We denote the probability mass in each control volume at time \(n\) by
\[
    q_i^n \;=\; \int_{\bm{X}_i} \rho^n(\bm{x}) \, d\bm{x}.
\]

Since \(\mathcal{P}\) acts linearly on \(\rho^n\), its discretization is a time-independent \emph{transition matrix} 
\(\bm{P} \equiv (P_{ij})\) 
that evolves the disctretized state-space vector \(\{q_j^n\}_{j=1}^{N}\) to \(\{q_i^{n+1}\}_{i=1}^{N}\) according to
\begin{equation}
\label{eq:trm}
    q_i^{n+1} 
    \;=\; \sum_{j=1}^{N} P_{ij}\,q_j^n.
\end{equation}
By construction, the transition matrix \(\bm{P}\) satisfies:
\[
    \sum_{i=1}^{N} P_{ij} \;=\; 1 
    \quad \text{for each } j.
\]
Moreover, since \(P_{ij}\) represents the probability of transition from cluster \(j\) to cluster \(i\), it also holds that
\begin{equation}
    0 \;\leq\; P_{ij} \;\leq\; 1.
    \label{eq:P_bound}
\end{equation}
When each control volume \(\bm{X}_j\) is mapped forward by the underlying flow, the probability mass is reallocated to possibly multiple volumes \(\bm{X}_i\). Thus, the entries of \(\bm{P}\) encode the local stretching or contraction of volumes under the flow map, which is directly linked to the \emph{Jacobian} of the forward map $\bm{f}$. As we refine the partition the transition matrix \(\bm{P}\) approaches the action of the continuous Perron--Frobenius operator, and each transition \(P_{ij}\) may be interpreted as a (coarse-grained) representation of volume expansion/contraction governed by the Jacobian. We present next an explicit derivation of how these matrix elements relate to the underlying Jacobian structure of the dynamical system.

Following Refs. \cite{cvitanovic2005chaos,beck1994thermodynamics}, we have 
\[
    \rho^{n+1}(\bm{x}) = \int_{\bm{X}} d\bm{y}\, 
    \delta(\bm{x}-\bm{f}(\bm{y})) \rho^{n}(\bm{y}) =
    \sum_{k=1}^{M} \frac{\rho^{n}(\bm{f}^{-1}_{k}(\bm{x}))}{|J(\bm{f}^{-1}_{k}(\bm{x}))|},
\]
for each $\bm{x} \in \Omega$, where $\bm{f}^{-1}_{k}$ denotes the $k \in \{1, ..., M \}$ preimages of the forward map $\bm{f}(\bm{x})$, and $J$ is the Jacobian of the transformation. Integrating both sides of the equation in state space over the cluster $\bm{X}_i$ yields 
\begin{equation}
    q_i^{n+1} = \int_{\bm{X}_i} \rho^{n+1}(\bm{x})\ d\bm{x}  
              = \int_{\bm{X}_i} \sum_{k=1}^{M} \frac{\rho^{n}(\bm{f}^{-1}_{k}(\bm{x}))}{|J(\bm{f}^{-1}_{k}(\bm{x}))|}\ d\bm{x}.
    \label{cons_prob_int}
\end{equation}
Now suppose that the partition $\bm{X}$ is sufficiently fine, and that the Jacobian is sufficiently smooth, so that in the integral of Eq.~(\ref{cons_prob_int}) it can be approximated by its value at the centroid $\bm{C}_i$ of $\bm{X}_i$. Under these assumptions, we obtain
\begin{equation}
\begin{split}
    q_i^{n+1} &\approx \sum_{k=1}^{M} \frac{1}{\bigl|J\bigl(\bm{f}^{-1}_{k}(\bm{C}_i)\bigr)\bigr|}\int_{\bm{X}_i}  \rho^{n}\bigl(\bm{f}^{-1}_{k}(\bm{x})\bigr)\ d\bm{x} \\
    &= \sum_{k=1}^{M} \frac{1}{\bigl|J\bigl(\bm{f}^{-1}_{k}(\bm{C}_i)\bigr)\bigr|}\,\sum_{j=1}^{N} p_{ij}^k\,q_j^n,
\end{split}
\label{cons_prob_int1}
\end{equation}
where in the second step we have defined 
\[
    \sum_{j=1}^{N} p_{ij}^k\,q_j^n \;:=\;\int_{\bm{X}_i}  \rho^{n}\bigl(\bm{f}^{-1}_{k}(\bm{x})\bigr)\ d\bm{x}\,.
\]
The coefficients $p_{ij}^k$ are the elements of a non-negative matrix representing the Koopman operator (the adjoint of the Perron--Frobenius operator) of $\bm{f}^{-1}_{k}$
\[
\rho^{n}\bigl(\bm{f}^{-1}_{k}(\bm{x})\bigr) = 
\int_{\bm{X}} d\bm{y}\, 
    \delta(\bm{y}-\bm{f}^{-1}_k(\bm{x})) \rho^{n+1}(\bm{y})
    \,,
\]
on a discretized state space.  The Koopman operator evolves the probability densities \emph{backward} in time through the inverse map $\bm{f}^{-1}_{k}$.  Concretely, each $p_{ij}^k$ encodes the probability that the $k$-th preimage of the $i$-th cluster centroid, i.e.\ $\bm{f}^{-1}_{k}(\bm{C}_i)$, lies in cluster $\bm{X}_j$.  In addition, the $p_{ij}^k$ obey the normalization conditions and bounds:
\begin{eqnarray}
\label{p_prop1}
& \sum_{j=1}^{N} p_{ij}^k \;=\; 1 \quad\text{for all}\ i,k,\\
\label{p_prop2}
& 0 \;\leq\; p_{ij}^k \;\leq\; 1 \quad\text{for all}\ i,j,k.
\end{eqnarray}
Rewriting Eq.~(\ref{cons_prob_int1}) as
\[
    \begin{split}
q^{n+1}_i &\approx \sum_{k=1}^{M} \frac{1}{|J(\bm{f}^{-1}_{k}(\bm{C}_i))|}\sum_{j=1}^{N} p_{ij}^k q^{n}_j\\&=\sum_{j=1}^{N}\left(\sum_{k=1}^{M} \frac{1}{|J(\bm{f}^{-1}_{k}(\bm{C}_i))|}p_{ij}^k \right)q^{n}_j
     =\sum_{j=1}^{N} P_{ij}q^{n}_j
\end{split}
\]
we obtain the relation between the deterministic forward map $\bm{f}$ and the transition matrix of Eq.~(\ref{eq:trm})
\[
    P_{ij} = \sum_{k=1}^{M} \frac{1}{|J(\bm{f}^{-1}_{k}(\bm{C}_i))|}p_{ij}^k
\,.
\]
If $M=1$ for all $i$, $p_{ij}$ coincides with the transpose of the adjoint of the transition matrix. In this case the Jacobian can be estimated directly from the transition matrix.
For $k>1$ we define
\begin{equation}
    A_i = \sum_{j=1}^{N} P_{ij} = \sum_{k=1}^{M} \frac{1}{|J(\bm{f}^{-1}_{k}(\bm{C}_i))|} \sum_{j=1}^{N} p_{ij}^k = \sum_{k=1}^{M} \frac{1}{|J(\bm{f}^{-1}_{k}(\bm{C}_i))|}
\,,
    \label{Aeq}
\end{equation}
where we used Eq.~(\ref{p_prop1}).

Let $k^*(i,j)$ be the index of the largest entry of $p_{ij}^k$ for given $i,j$. We can then write:
\[
\begin{split}
    B_j &=  \max_i P_{ij} =  \max_i\sum_{k=1}^{M}  \frac{1}{|J(\bm{f}^{-1}_{k}(\bm{C}_i))|} p_{ij}^k \\&\approx  \max_i\sum_{k=1}^{M}  \frac{1}{|J(\bm{C}_j)|} p_{ij}^k\approx \frac{1}{|J(\bm{C}_j)|}\max_i p_{ij}^{k^*(i,j)} \leq \frac{1}{|J(\bm{C}_j)|}
\,,
\end{split}
\]
where we approximated with $\bm{C}_j$ all the preimages of $\bm{C}_i$ falling in $\bm{X}_j$ and we used Eq.~(\ref{p_prop2}).
Furthermore, since $B_j$ is bounded by probability bound Eq.~(\ref{eq:P_bound}), we have that
\begin{align}
    B_j  = \max_i P_{ij} \leq  \text{min} \{ |J(\bm{C}_j)|^{-1} , 1 \}
\,.  
    \label{Beq2}
\end{align}
For an alternative derivation of the expressions for $A_i$ and $B_j$, see the Appendix.

The Perron--Frobenius operator \(\mathcal{P}\) is often characterized as a contraction, satisfying \(\|\mathcal{P}\rho\|\le \|\rho\|\). Thus, \(A_i\) defined in Eq.~(\ref{Aeq}) can be viewed as the local inverse dissipation rate of the state-space volume centered at \(\bm{C}_i\). In one dimension $B_j$ defined in Eq.~(\ref{Beq2}) is an upper bound on the derivative of the forward map.

An addition of noise to each $\bm{x}$ does not alter $A_i$ (Eq.~\ref{Aeq}) because each $p_{ij}^k$ remains normalized. However, noise increases the variance of $p_{ij}^k$, thereby reducing its maximal values. In regions where $\bm{f}^{-1}$ contracts the state space, $p_{ij}^k\approx 1$ are weakly affected if the noise remains small compared to the control volumes. Consequently, certain entries $B_j$ (Eq.~\ref{Beq2}) will stay close to $1/|J(\bm{C}_j)|$, allowing for the extraction of the Jacobian.

The expressions for \(A_i\) and \(B_j\) are based on the discretetization of matrices \(P_{ij}\). Although we shall rely on explicit formulas for the Jacobian and the pre-image maps \(\bm{f}_{k}^{-1}\) in our examples, we emphasize that the discrete transfer operator may be inferred purely from data, without requiring prior knowledge of the governing equations.

\section{Jacobian of chaotic maps}
\label{sec3}

In this section, we apply the method described above to a variety of one-dimensional and two-dimensional dynamical systems. Our primary objective is to relate the entries of the numerically estimated transfer operator to the Jacobian of the underlying system. 

\subsection{One-dimensional chaotic maps}

For detailed discussions of the dynamical systems that follow, the reader is referred to Chapter 17 of \cite{beck1994thermodynamics}.

\subsubsection{The Ulam Map}
The mapping 
\[
    x_{n+1} = 1 - 2(x_n)^2,
\]
on the interval $x_n \in (-1,1]$ for all $n$ is known as `Ulam map'. 
The Perron-Frobenius evolution equation for the probability density is 
\[
    \rho^{n+1}(C_i) = \frac{1}{4}\left(\frac{2}{1-C_i}\right)^{\frac{1}{2}} \left[\rho^n\left(\left(\frac{1-C_i}{2}\right)^{\frac{1}{2}}\right)+\rho^n\left(-\left(\frac{1-C_i}{2}\right)^{\frac{1}{2}}\right) \right],
\]
and, subsequently, 
\begin{equation}
    \sum_k \frac{1}{|J(f^{-1}_{k}(C_i))|} = \frac{\sqrt{2}}{2} \left(\frac{1}{1-C_i}\right)^{\frac{1}{2}},
    \label{A_num1}
\end{equation}
with
\begin{equation}
    \frac{1}{|J(C_j)|} = \frac{1}{4|C_j|}
\,.
    \label{B_num1} 
\end{equation}

\subsubsection{Continued Fraction Map}

In the case of the continued fraction map, given by 
\[
    \bm{f}(x) = \frac{1}{x}-\left\lfloor\frac{1}{x} \right\rfloor
\,,\qquad
    x_n \in [0,1]\;\;\; \forall n
\,,
\]
we have 
\[
    \rho^{n+1}(C_i) = \sum_{k=1}^\infty\frac{1}{(k+C_i)^2} \rho^n\left(\frac{1}{C_i+k}\right),
\]
\begin{equation}
    \sum_k \frac{1}{|J(f^{-1}_{k}(C_i))|} = 
    \Psi(1+C_i),
    \label{A_num2}
\end{equation}
and
\begin{equation}
    \frac{1}{|J(C_j)|} = C_j^2,
\label{B_num2}
\end{equation}
where $\Psi(x)=
\frac{\Gamma'(x)}{\Gamma(x)}$ is the polygamma function (logarithmic derivative of the Gamma function), and $\left\lfloor x \right\rfloor$ denotes the integer part of $x.$

\subsubsection{Cusp Map}

The map
\[
    \bm{f}(x) = 1 - 2 |x|^{\frac{1}{2}}
\,,\qquad 
x_n \in [-1,1]\;\;\; \forall n
\]
is known as the `cusp map', with the Perron-Frobenius evolution equation for the probability density 
\[
    \rho^{n+1}(C_i) = \frac{1-C_i}{2} \left[\rho^n\left(\frac{(1-C_i)^2}{4}\right)+\rho^n\left(-\frac{(1-C_i)^2}{2}\right)\right],
\]
\begin{equation}
    \sum_k \frac{1}{|J(f^{-1}_{k}(C_i))|} = 1+C_i,
    \label{A_num3}
\end{equation}
and
\begin{equation}
    \frac{1}{|J(C_j)|} = \sqrt{|C_j|}.
    \label{B_num3}
\end{equation}

\subsubsection{One-dimensional Chebyshev map}

The one-dimensional Chebyshev map is defined as:
\begin{equation}
    x_{n+1} = T_N(x_n)
\,,\qquad
    x \in [-1,1]
\,,
    \label{eq:tcheby1d_map}
\end{equation}
where $T_N(x)$ is the $N$-th Chebyshev polynomial
\(
    T_N(x) = \cos(N \arccos(x))
\,.
\)
The Perron-Frobenius evolution equation for this map is
\begin{equation}
    \sum_{k=0}^{N-1} \frac{1}{|J(f^{-1}_k(C_i))|} = \sum_{k=0}^{N-1} \frac{\sqrt{1 - x_k^2(C_i)}}{N \left| \sin(N \arccos(x_k(C_i))) \right|}
\,,
    \label{eq:tcheby1d_A}
\end{equation}
and
\begin{equation}
    \frac{1}{|J(C_j)|} = \frac{\sqrt{1 - C_j^2}}{N \left| \sin(N \arccos(C_j)) \right|}
\,,
    \label{eq:tcheby1d_jacobian}
\end{equation}
where
\[
    x_k(C_i) = \cos\left( \frac{\arccos(C_i) + 2\pi k}{N} \right), \quad k = 0, 1, \dots, N-1
\,.
\]

\subsubsection{Results}

We partition the state space into ${N}=100$ equally sized control volumes, and assign every orbit point to its corresponding volume. Using this cluster of trajectories, we construct the transition matrix $\bm{P}$, from which we extract our estimates for the inverse dissipation $A_i=\sum_{j=1}^{N} P_{ij}$ and the local instability rate $B_j=\mathrm{max}_i P_{ij}$ from the data collected by running the dynamics. We also repeat the procedure with additive Gaussian white noise of amplitude $\sigma$.

\begin{figure*}
\centering
\includegraphics[width=0.7\linewidth]{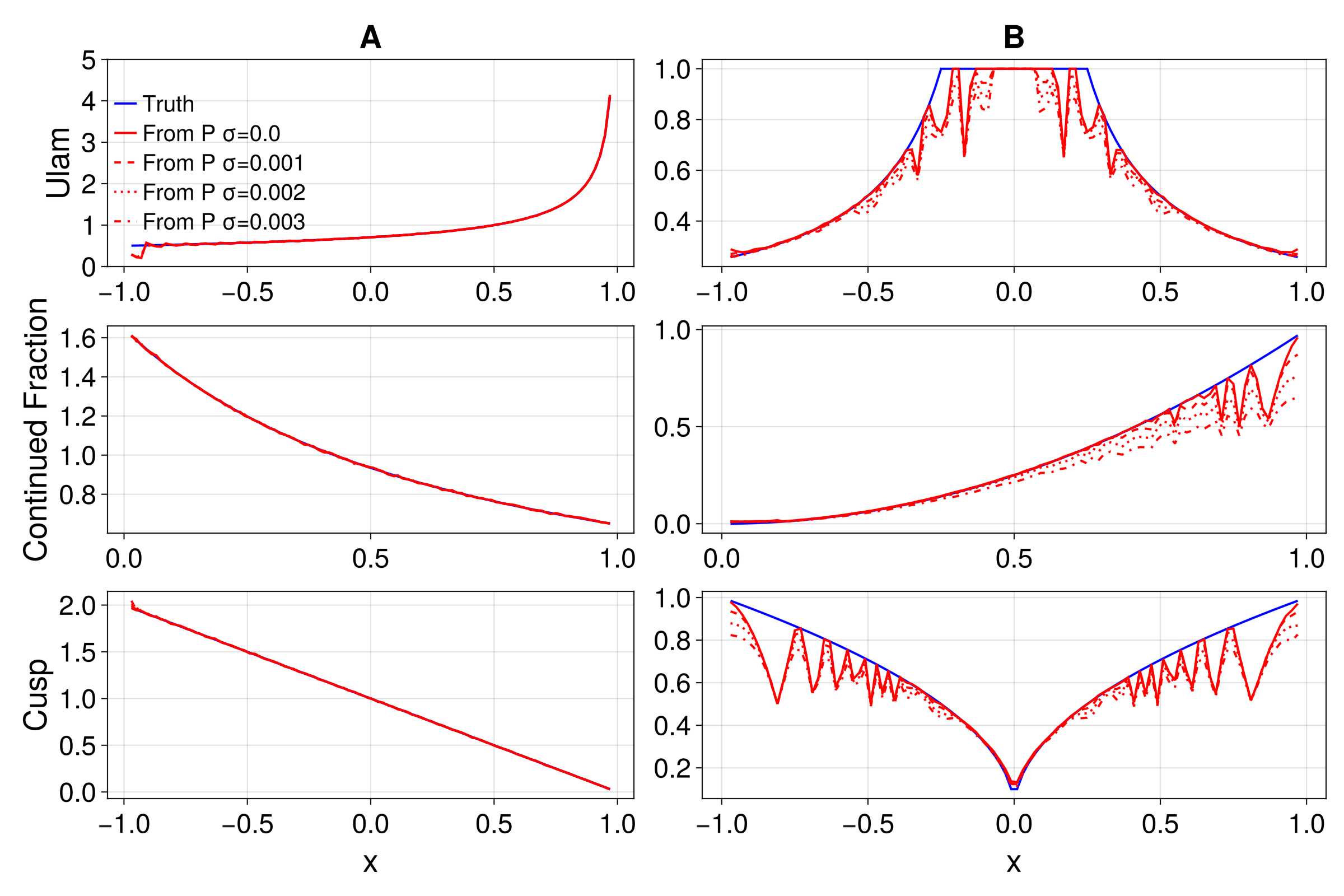}
\caption{
\textit{Ulam, continued fraction and cusp maps:} Plots of $A_i$ and $B_j$ (left and right column, respectively) derived from the transition matrix (depicted with red lines), along with their analytical estimates (shown in blue), for the Ulam, continued fraction, and cusp maps. Noise amplitudes of $\sigma = 0, 0.001, 0.002, 0.003$ have been applied to the maps.}
    \label{Fig1}            
\end{figure*}

\begin{figure*}[ht]
    \centering    \includegraphics[width=0.9\linewidth]{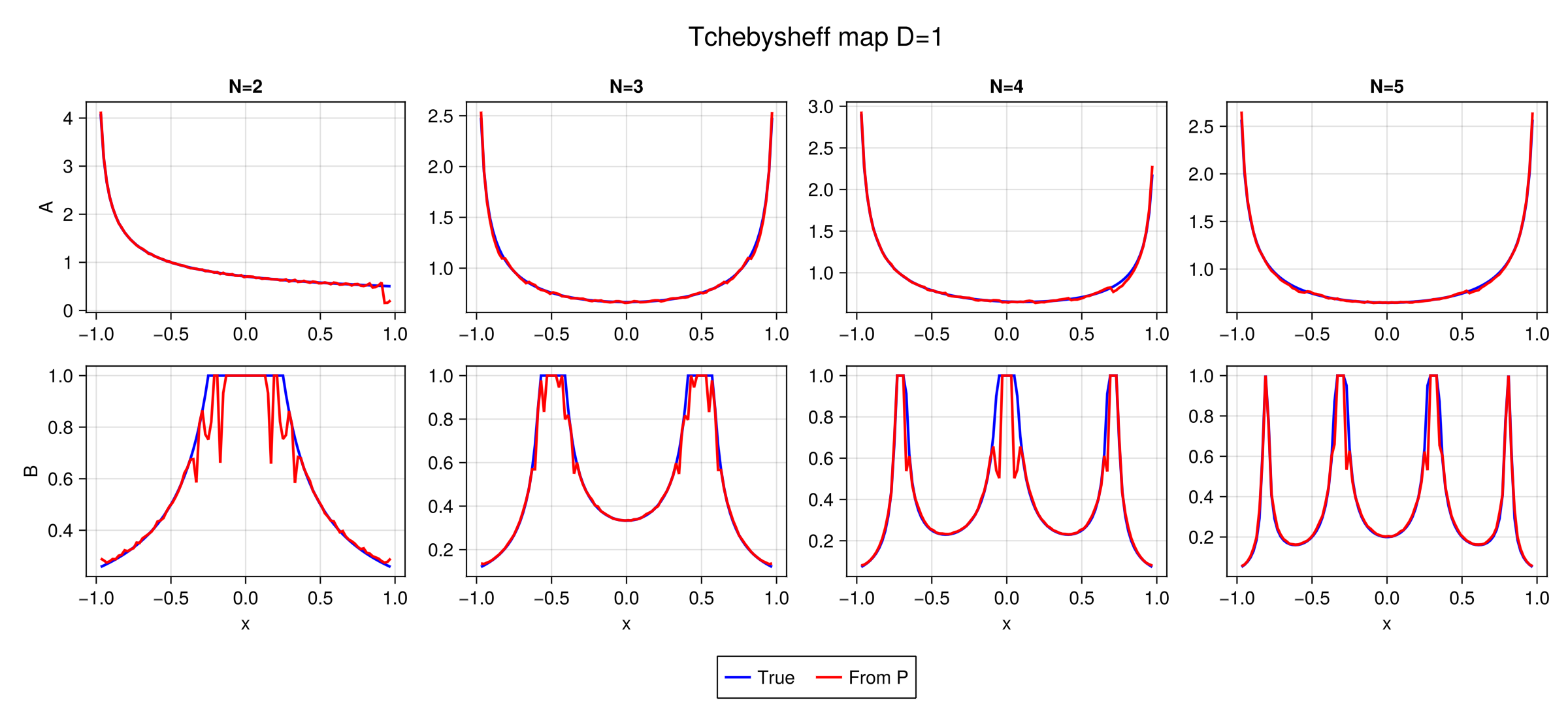}
    \caption{\textit{One-dimensional Chebyshev map:} Plots of $A_i$ and $B_j$ derived from the transition matrix (red lines), along with their analytical estimates (blue lines), for the Chebyshev map defined in Eq.~(\ref{eq:tcheby1d_map}) with $N=2, 3, 4$, and $5$, and $\sigma=0$.}
    \label{fig:tchebysheff_1d}
\end{figure*}

In Figs.~(\ref{Fig1}, \ref{fig:tchebysheff_1d}) we compare
the expressions $A_i$ and $B_j$ obtained from the transition matrix with their analytical values, as defined in Eqs.~(\ref{A_num1}) to 
(\ref{eq:tcheby1d_A}). In the panels representing $B_j$ we set to unity all values of the analytical estimate of $B_j$ larger than one, since the elements of $B_j$ obtained from the transition matrix cannot be larger than unity.

Upon varying the noise amplitudes in Fig.~(\ref{Fig1}), we observe a consistent pattern: the `data' estimate of $A_i$ accurately reproduces its expected value, without appreciable deviations, up to the precision of our numerics, with or without additive noise. On the other hand, the data estimate of $B_j$ is less accurate, as well as more sensitive to noise. In particular, the local instability rate is often underestimated as the noise amplitude increases.
However, the quality of the estimates of $B_j$ does follow a pattern that depends on the instability itself: 
it is in fact noted from the plots that the estimate for the instability rate $B_j$ is consistently the more accurate and noise-robust, the larger the Jacobian, or equivalently, the smaller $B_j$.     

\subsection{Two-Dimensional Coupled Chebyshev Maps}

This section demonstrates the application of the proposed methodology to estimate the Jacobian for two-dimensional coupled Chebyshev maps \cite{dettmann2005periodic}. We will compute coefficients $A_i$ and $B_j$ for small coupling strengths and polynomial orders $N=2,3,4$.

The two-dimensional coupled Chebyshev map is defined by:
\begin{equation}
\begin{aligned}
    x_{n+1} &= (1-a)\,T_N(x_n) + a\,T_N(y_n), \\
    y_{n+1} &= (1-a)\,T_N(y_n) + a\,T_N(x_n),
\end{aligned}
\label{eq:tcheby2d_map}
\end{equation}
where $T_N(\cdot)$ is the $N$-th Chebyshev polynomial, and the parameter $a \in [0,1]$ controls the coupling strength. 
For the particular cases $N=2,3,4$, the polynomials are
\begin{itemize}
    \item $T_2(x) = 2x^2 - 1,$
    \item $T_3(x) = 4x^3 - 3x,$
    \item $T_4(x) = 8x^4 - 8x^2 + 1.$
\end{itemize}
Coupled Chebyshev maps with a relatively weak coupling 
(here the parameter $a$ is set to $0.01$ throughout the simulations presented) are ergodic and mixing like their uncoupled counterparts, yet they exhibit multidimensional features (e.g. their natural measure), whereas they tend to develop coherent structures and synchronization for stronger couplings~\cite{yanbeck22}. Moreover, still in the regime of weak coupling, the dynamics is locally expanding (both eigendirections are unstable), rather than hyperbolic (only one is unstable)~\cite{dettmann2005periodic}, which yields non-trivial dissipation and instability fields in the state space. For that reason in particular, the local Jacobian is still a measure of instability, and so is our observable $B_j$.      

The Jacobian of the forward map in Eq.~\eqref{eq:tcheby2d_map} is:
\[
\begin{split}
    |J(x,y)| \;=\; 
    \Bigl|
       (1 - a)^2\,T_N'(x)\,T_N'(y) 
       \;-\; a^2\,T_N'(y)\,T_N'(x)
    \Bigr|,
\end{split}
\]
where
\[
    T_N'(x) \;=\; \frac{-N\,\sin\bigl(N \,\arccos(x)\bigr)}{\sqrt{1 - x_2}}.
\]
Although the inverse of the two-dimensional map does not admit a closed-form solution for arbitrary $N$, it can be determined  analytically for $N=2$: 
\[
    x_n \;=\; \pm \, \sqrt{\frac{-1 - x_{n+1} + a \left(2 + x_{n+1} + y_{n+1}\right)}{-2 + 4a}},
\]
\[
    y_n \;=\; \pm\,\sqrt{\frac{-1 - y_{n+1} + a \left(2 + x_{n+1} + y_{n+1}\right)}{-2 + 4a}},
\]
and $N=4$:
\[
    x_n \;=\; \pm\,\frac{1}{2} \sqrt{2 \pm \sqrt{\frac{-2 \left(1 + x_{n+1}\right) + 2a \left(2 + x_{n+1} + y_{n+1}\right)}{-1 + 2a}}}\,,
\]
\[
    y_n \;=\; \pm \,\frac{1}{2} \sqrt{2 - \sqrt{\frac{-2 \left(1 + y_{n+1}\right) + 2a \left(2 + x_{n+1} + y_{n+1}\right)}{-1 + 2a}}}.
\]
For $T_3$, the inverse map is given by the solution to the cubic equations:
\[
    (x_n)^3 \;-\;\tfrac{3}{4}\,x_n \;+\;\tfrac{a}{4(1 - 2a)}\,y_{n+1}
    \;+\;\tfrac{a - 1}{4(1 - 2a)}\,x_{n+1} \;=\; 0,
\]
\[
    (y_n)^3 \;-\;\tfrac{3}{4}\,y_n \;+\;\tfrac{a}{4(1 - 2a)}\,x_{n+1}
    \;+\;\tfrac{a - 1}{4(1 - 2a)}\,y_{n+1} \;=\; 0.
\]

\begin{figure*}[ht]
    \centering
    \includegraphics[width=0.8\linewidth]{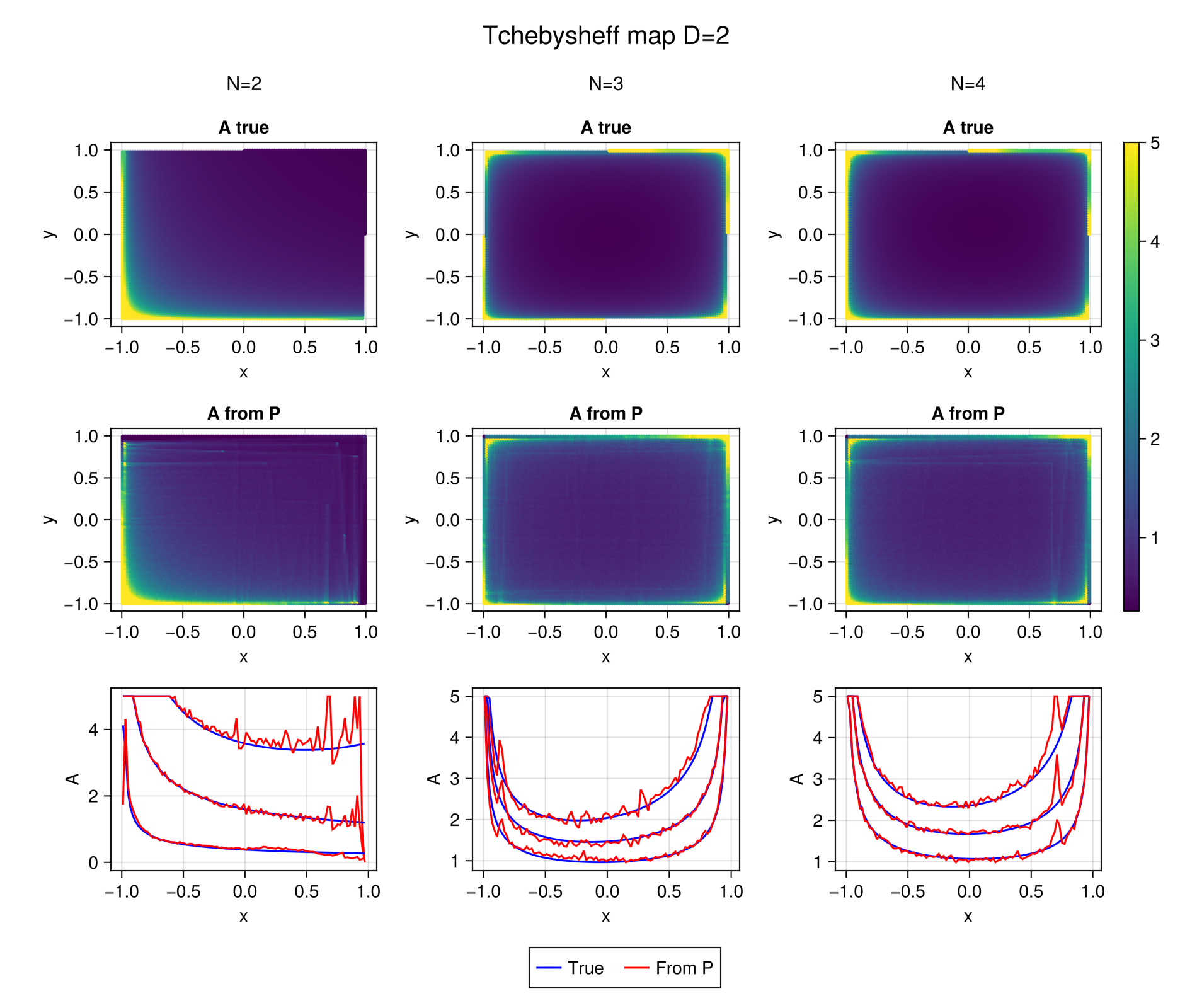}
    \caption{\textit{Two-dimensional coupled Chebyshev map.} Heat maps and scatter plots comparing the $A_i$ values from the transition matrix with their analytical counterparts for the map in Eq.~\eqref{eq:tcheby2d_map}, using $N=2,3,4$ and coupling parameter $a = 0.01$. Below the heatmaps, we compare $A_i$ as a function of the $x$ coordinate for different values of $y$ ($y=-0.98, -0.9, 0.8$ for $N=2$ and $y=-0.98, -0.96, 0.8$ for $N=3,4$), obtained through the two different methods. 
    }
    \label{fig:tchebyshev_2d_A}
\end{figure*}

\begin{figure*}[ht]
    \centering
    \includegraphics[width=0.8\linewidth]{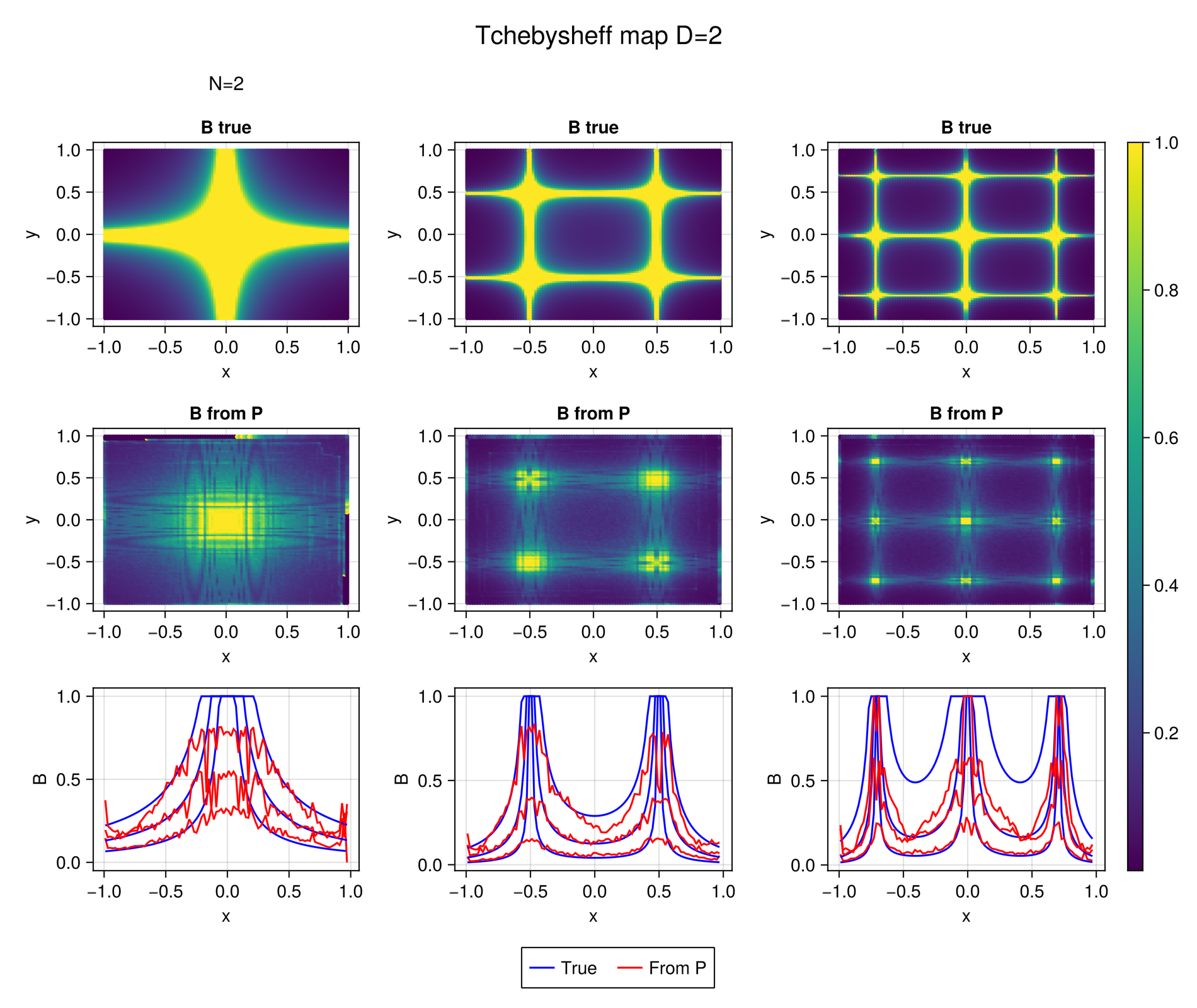}
    \caption{\textit{Two-dimensional coupled Chebyshev map.} Same as Fig. (\ref{fig:tchebyshev_2d_A}), but for $B_j$. The values of $y$ used in the plots in the last row are $y=-0.98, -0.5, -0.3$ for $N=2$, $y=-0.98, -0.7, -0.4$ for $N=3$ and $y=-0.4, -0.1, -0.04$ for $N=4$. 
    }
    \label{fig:tchebyshev_2d_B}
\end{figure*}

We simulate the Chebyshev map over $10^7$ time steps for $N=2, 3, 4$, partitioning the state space into a $100 \times 100$ grid of equally size control volumes to construct the transition matrix. Figs.~(\ref{fig:tchebyshev_2d_A},
\ref{fig:tchebyshev_2d_B}) present the results for $N=2,3,4$, comparing the numerical estimates of $A_i$, $B_j$ with their analytical counterparts. Similar observations to those made in previous examples apply here. Specifically, we observe that $A_i$ closely matches its analytical expectation across the state space. In contrast, the observable $B_j$ computed from data is accurate primarily in the more unstable regions of the state space, while it provides poorer estimates of the inverse of the Jacobian in the dynamically less unstable or nearly marginal regions.

\section{Conclusions}
\label{conclusions}

In this study we have developed and validated a novel methodology that connects the dissipation and instability fields to the transition matrix of a dynamical system, constructed by partitioning state space into a finite set of control volumes. 
The task is made possible by inferring constraints on the Jacobian from the statistical features of the dynamical system, which can be easily obtained from observables without any {\it{a priori}} assumptions on the governing equations. Knowledge of the Jacobian is pivotal in deriving pertinent information about the dynamical features of the system, and it paves the way for the development of accurate prediction or control tools.

The numerical simulations of both one-dimensional and two-dimensional chaotic maps demonstrate the robustness of our methodology -- but also highlight specific challenges. In particular, in the two-dimensional case, we observe that state-space regions, where the local Jacobian is accurately reproduced by the data-determined observable $B,$ are bounded.

One potential approach we aim to explore in the future involves synthesizing the less accurate information from the estimate $B$ for the local Jacobian, with the more precise data from the observable $A$ for the dissipation. This strategy would involve initially using $B$ to aid in the estimation of the deterministic flow/map $\bm{f}$. Subsequently, the inverse function $\bm{f}^{-1}$ could be calculated, allowing us to verify the accuracy of our estimation. This verification would be accomplished by using $\bm{f}^{-1}$ to estimate the dissipation $A$, followed by a matching of this result to $A$, this time as derived from the transition matrix via Eq.~(\ref{Aeq}). This hybridized approach may help overcome the limitations associated with using the observable $B$ alone for the reconstruction of the Jacobian.

Another challenge can arise from errors in determining the entries of the transition matrix corresponding to regions of coordinate space that are rarely visited by the dynamical system, a common occurrence when only a limited amount of data is available. This introduces noise into both observables $A$ and $B$. The resulting effects are more pronounced in $B$, since each entry corresponds directly to an individual entry from the transition matrix, unlike $A$, where each entry is a sum of all the elements in each column, effectively averaging out the noise introduced by the finite amount of processed data.

In summary, the methodology developed in this study provides a new pathway to connect statistical features of dynamical systems, specifically the transition matrix, with the local dynamics captured by the Jacobian. While our approach shows promise in effectively reconstructing the Jacobian from observed data, it also highlights challenges, particularly in higher-dimensional systems where the relationship between the transition matrix and the Jacobian becomes less straightforward. Future research will focus on refining the approach by combining the insights from both observables $A$ and $B$ to improve the accuracy and reliability of the Jacobian reconstruction. Additionally, expanding this methodology to handle more complex dynamical systems and exploring the effects of noise and other perturbations will be crucial for broadening the applicability of this approach and for demonstrating its potential. 

\section*{Acknowledgements}
The authors want to thank the 2022 Geophysical Fluid Dynamics Program, where a significant portion of this research was undertaken; the GFD Program is supported by the National Science Foundation Grant No. 1829864, United States and the Office of Naval Research, United States.
We are grateful to Edward A. Spiegel for his unwavering support through the Woods Hole GFD turbulent summers that begot this collaboration.
LTG was supported by the Swedish Research Council (Vetenskapsradet) Grant No. 638-2013-9243. AS acknowledges support by Schmidt Sciences, LLC, through the Bringing Computation to the Climate Challenge, an MIT Climate Grand Challenge Project. 

\bibliography{references}

\section*{Appendix: Alternative derivation of the expressions for $A_i$ and $B_j$}
Let us write a density in the state space as a sum of contributions over the discretization 
Eq.~(\ref{contrVol}),
$\bm{X}=[\bm{X}_1,\dots,\bm{X}_N]$,
\begin{equation}
\rho^n(\bm{x}) = \sum_j \alpha_j^n\rho_j(\bm{x})
\,,
\label{dens_part}
\end{equation}
where the $\rho_j$'s are smooth functions peaked at the centroid $\bm{C}_j$ of each interval, for example the Gaussians
\[
\rho_j(\bm{x}) = \frac{1}{(2\pi\sigma^2)^{D/2}} \exp\left( -\frac{\|\bm{x} - \bm{C}_j\|^2}{2\sigma^2} \right)
\,.
\]
We set $\sigma^2 = [\Omega / {N}]^{2/D}/2\pi$ to ensure that the support of $\rho_j$ approximately coincides with $\bm{X}_j$, thereby rendering the Gaussian basis nearly orthogonal. Consequently, we can write:
\[
q^n_i = \int_{\bm{X}_i} \rho^n(\bm{x}) \, d\bm{x} \approx \alpha_i^n.
\]
The evolution of the probability of each interval reads
\[
q^{n+1}_i = \int_{\bm{X}_i} \rho^{n+1}(\bm{x})d\bm{x} = \int_{\bm{X}_i} d\bm{x} \int_{\bm{X}} d\bm{y}\, \delta(\bm{x}-\bm{f}(\bm{y})) \rho^n(\bm{y}) \,   
\]
where the Perron-Frobenius operator $\int_{\bm{X}} dy\, \delta(x-\bm{f}(y)) \cdot$  transports state-space densities forward in time. Using the partition~\eqref{dens_part}, the previous is rewritten as
\begin{eqnarray}
q^{n+1}_i &=& \sum_{j=1}^{N}\alpha_j^n \int_{\bm{X}_i} d\bm{x}  \int_{\bm{X}} d\bm{y} \delta(\bm{x}-\bm{f}(\bm{y}))\rho_j(\bm{x}) \\
&\approx& \sum_{j=1}^{N} q_j^n \int_{\bm{X}_i} d\bm{x}  \int_{\bm{X}_j} d\bm{y} \delta(\bm{x}-\bm{f}(\bm{y}))\rho_j(\bm{x}) 
    \nonumber\\ 
          &=&
\sum_{j=1}^{N}q_j^n\sum_{k=1}^{M} \int_{\bm{X}_i} d\bm{x} 
\frac{\rho_j(\bm{f}^{-1}_{k}(\bm{x}))}{|J(\bm{f}^{-1}_{k}(\bm{x}))|}
\,.
    \nonumber
\end{eqnarray}

Now suppose that the partition $\bm{X}$ is sufficiently fine for the integrals to be well approximated by the value of the integrand at the centroid $\bm{C}_i$ of $\bm{X}_i$ times the measure of the interval.
Recalling that the $\rho_j$'s are Gaussians, we write
\begin{equation}
q^{n+1}_i \simeq \sum_{j=1}^{N} q_j^n \sum_{k=1}^{M} \frac{e^{-\|\bm{f}^{-1}_{k}(\bm{C}_i)-\bm{C}_j\|^2/2\sigma^2}}{|J(\bm{f}^{-1}_{k}(\bm{C}_i))|} 
\,,
\label{qiapprox}
\end{equation}
having chosen $\sigma^2=[\mu(\bm{X}_i)]^{2/D}/2\pi$ for 
the normalization to be unity. One can then approximately identify the entries of the transition matrix with
\[
P_{ij} \simeq \sum_{k=1}^M \frac{e^{-\|\bm{f}^{-1}_{k}(\bm{C}_i)-\bm{C}_j\|^2/2\sigma^2}}{|J(\bm{f}^{-1}_{k}(\bm{C}_i))|} 
\,.    
\]
The probabilities $P_{ij}$ to go from $\bm{X}_j$ to $\bm{X}_i$ in one iteration of the transfer operator are constrained by the condition
\[
\sum P_{ij} =1
\,,
\]
so that 
\[
\sum_{i=1}^{N}\sum_{k=1}^M \frac{e^{-\|\bm{f}^{-1}_{k}(\bm{C}_i)-\bm{C}_j\|^2/2\sigma^2}}{|J(\bm{f}^{-1}_{k}(\bm{C}_i))|} \simeq 1 
\,,
\]
and, as every single $i-$th contribution is positive definite, we have 
\[
 \sum_{k=1}^M \frac{e^{-\|\bm{f}^{-1}_{k}(\bm{C}_i)-\bm{C}_j\|^2/2\sigma^2}}{|J(\bm{f}^{-1}_{k}(\bm{C}_i))|} {\lesssim} 1
 \,.
\]
The largest entries of the transfer matrix are realized when $\bm{X}_j$ maps into $\bm{X}_i$, that is for a special $k^*$ such that $\|\bm{f}^{-1}_{k^*}(\bm{C}_i)-\bm{C}_j\|^2<\sigma^2$, and so we have rederived  Eq.~(\ref{Beq2}),
\begin{equation}
B_j \equiv \max_i P_{ij} \simeq  \frac{1}{|J(\bm{f}^{-1}_{k^*}(\bm{C}_i))|} \approx \frac{1}{|J(\bm{C}_j)|}
\,.
\label{Beq}
\end{equation}
All the previous expressions from~\eqref{qiapprox} up to~\eqref{Beq} rely on the assumption of non-vanishing Jacobians $|J(\bm{f}^{-1}_{k^*}(\bm{C}_i))|$ and
$|J(\bm{C}_j)|$. If, instead, the Jacobian does vanish somewhere in the intervals $\bm{X}_i$ or
$\bm{X}_j$, the approximation~\eqref{qiapprox} no longer holds, and
\[
 \int_{\bm{X}_i} d\bm{x} 
\frac{\rho^{n}_j(\bm{f}^{-1}_{k}(\bm{x}))}{|J(\bm{f}^{-1}_{k}(\bm{x}))|} =  \int_{\bm{f}_k^{-1}(\bm{X}_i)} d\bm{y}\, \rho^n_j(\bm{y}) \] 
\[= \int_{\bm{f}_k^{-1}(\bm{X}_i)}  \frac{d\bm{y}}{(2\pi\sigma^2)^{D/2}} e^{-\|\bm{y}-\bm{C}_j\|^2/2\sigma^2} \leq 1
\]
now provides an upper bound for $\max_i P_{ij}$. On the other hand,
the sum of the probabilities of landing in $\bm{X}_i$ is, confirming  Eq.~(\ref{Aeq}), 
\[
A_i \equiv \sum_j P_{ij} \simeq \sum_{j=1}^{N}\sum_{k=1}^M \frac{e^{-\|\bm{f}^{-1}_{k}(\bm{C}_i)-\bm{C}_j\|^2/2\sigma^2}}{|J(\bm{f}^{-1}_{k}(\bm{C}_i))|} \approx 
\sum_{k=1}^M \frac{1}{|J(\bm{f}^{-1}_{k}(\bm{C}_i))|}
\,,   
\]
estimating that the only non-negligible contributions to the sum over $j$ are coming from the $\bm{C}_j$'s such that $||\bm{f}^{-1}_{k}(\bm{C}_i)-\bm{C}_j||<\sigma^2$.

\end{document}